\documentclass[%
reprint,  
amsmath,amssymb,
aps,
prl,
]{revtex4-2}

\usepackage[pdftex]{graphicx}
\usepackage{bm}

\usepackage{mathtools}
\usepackage{physics2}
\usephysicsmodule{ab, ab.legacy, op.legacy}
\usepackage{fixdif, derivative}
\usepackage{booktabs}
\usepackage{comment}
\usepackage[dvipsnames]{xcolor}

\newcommand{\rmsub}[2]{#1_\mathrm{#2}}
\newcommand{\panlab}[1]{\textbf{(#1)}}

\newcommand{\jointarticle}{\cite{ishii2026Degree}}

\renewcommand{\section}[1]{\paragraph{#1.}}

\begin{document}


\title{%
Qualitatively distinct mechanisms of noise-induced escape in fully connected populations of diffusively coupled bistable elements
}

\author{Hidemasa Ishii}
\email{hidemasaishii1997@g.ecc.u-tokyo.ac.jp}
\author{Hiroshi Kori}%
\affiliation{%
Department of Complexity Science and Engineering, Graduate School of Frontier Sciences, the University of Tokyo, Chiba 277-8561, Japan
}%


\date{\today}

\begin{abstract}
The analysis of noise-induced escape in populations of bistable elements is challenging, because nonlinearity, coupling, and noise all play essential roles.
We show that the interplay of these three factors yields three qualitatively distinct escape mechanisms depending on coupling strength in populations of diffusively coupled bistable elements. 
To clarify dominant driving factors of escape dynamics, we develop a model-reduction approach, deriving three effective one-dimensional dynamics:
nonlinear mean-field Fokker-Planck equation in the weak-coupling regime, stochastic mean-field dynamics in the strong-coupling regime, and deterministic mean-field dynamics in the intermediate regime. 
We validate these reduced descriptions by comparing predicted mean escape times with numerical simulations. 
We identify a distinct dominant driving factor of collective escape in each regime. 
Notably, the three mechanisms emerge through the interplay of nonlinearity, diffusive coupling, and dynamical noise --- rather than bifurcations of the noise-free system. 
Our approach serves as a framework applicable to other stochastic nonlinear systems with diffusive coupling, motivating a further search for similar synergistic phenomena.
\end{abstract}

\maketitle


\section{Introduction}
External forcings on bistable systems induce switching between the two typical states~\cite{gammaitoni1998Stochastic,ashwin2012Tipping}.
This has motivated the use of bistable models in practical research on abrupt changes in systems's states, from epilepsy~\cite{suffczynski2004Dynamics,creaser2020Dominolike} and climate change~\cite{wunderling2021Interacting} to social uprisings~\cite{brummitt2015Coupled}.
When external forcings are stochastic, the switching behavior is known as
noise-induced escape~\cite{ashwin2017Fast}.
Populations of fluctuating bistable elements also exhibit noise-induced escape.
This phenomenon is theoretically interesting, because nonlinearity, interaction, and noise all play essential roles in the escape dynamics.

Previous studies~\cite{frankowicz1982Stochastic,ashwin2017Fast,creaser2018Sequential,ishii2024Diffusivea} have considered diffusively coupled populations of bistable elements, examining the influence of varying coupling strength on escape dynamics, often characterized by mean escape times.
However, existing analyses mostly build on the bifurcation structure of the noise-free system, focusing on small systems and limited parameter regimes.
Such a bifurcation-based approach is infeasible for large or networked populations, nor can it account for the non‑monotonic dependence of escape times on coupling strength observed in numerical studies~\cite{frankowicz1982Stochastic,ishii2024Diffusivea}.
Accordingly, a more coarse-grained theoretical approach is needed to understand how the interplay of nonlinearity, coupling, and noise shapes collective dynamics of noise-induced escape.

In this letter, we analyze large populations of globally coupled bistable elements.
We derive effective one-dimensional dynamics that approximate collective escape processes in weak-coupling, intermediate, and strong-coupling regimes, which is our main result.
Each of the reduced descriptions represents a distinct escape mechanism.
Our theoretical results are verified by direct numerical simulations of the original model.
We conclude this letter by highlighting the synergistic role of nonlinearity, diffusive coupling, and dynamical noise in the collective escape.
The findings are generalized to a class of networked systems in the companion article~\jointarticle{}.

\section{Model and mean escape time}

\begin{figure}[t]
  \centering
  \includegraphics[width=\linewidth]{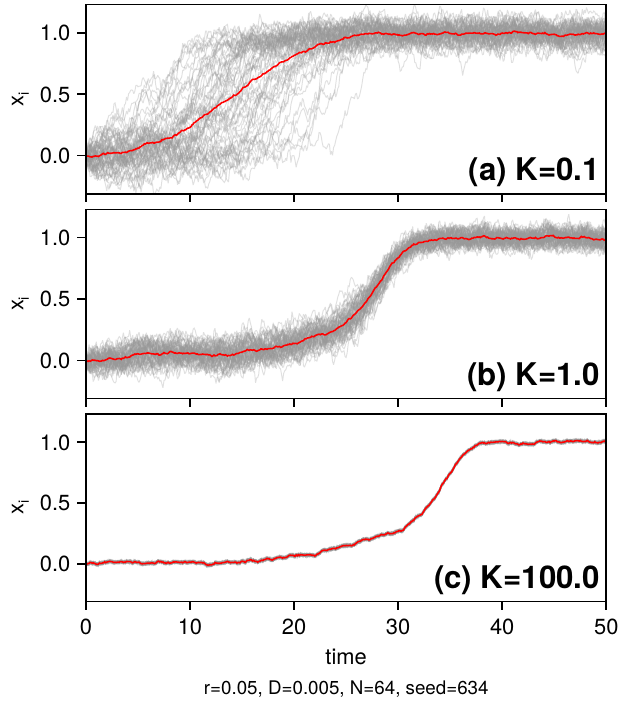}
  \caption{Typical trajectories in different regimes.
    Light gray lines depict the evolution of all the elements.
    Bright red lines show trajectories of the mean field.
    \panlab{a} Elements are non-synchronous in the weak-coupling regime.
    \panlab{b} In the intermediate regime, elements' states evolve around the mean field while maintaining non-negligible variance.
    \panlab{c} The population behaves as a single unit under strong diffusive coupling.
  }
  \label{fig:trajectories}
\end{figure}

We analyze the model governed by the following stochastic differential equations (SDEs):
\begin{subequations} \label{eq:model.original}
  \begin{align}
  \dot{x}_i =& f(x_i) + \frac{K}{N} \sum_{j=1}^N \pab{x_j - x_i} + \sqrt{2D} \xi_i,  \\
  f(x) \coloneq& -x \pab{x - r} \pab{x - 1},
\end{align} \end{subequations}
where $K$ is the coupling strength, $N$ is the system size, $D$ is the noise strength, and $\xi_i$ denotes pairwise independent white Gaussian noise satisfying $\aab{\xi_i(t)} = 0$ and $\aab{\xi_i(t) \xi_j(t+\tau)} = \delta_{ij} \delta\pab{\tau}$.
The local flow $f(x)$ induces bistability of each element in the absence of coupling and noise.
The system $\dot{x} = f(x)$ has stable fixed points at $x = 0$ and $1$ and an unstable fixed point at $x = r$.
Throughout this research, we assume $0 < r \ll 0.5$, in which case $x = 1$ is much more stable than $x = 0$.
In the following, we refer to $x = 0$ and $x = 1$ as the ``background'' and ``active'' states, respectively.
The strong asymmetry in bistability allows us to focus on the escape from the background to active states, ignoring the probability of the opposite transition.
While a small $r$ is crucial for numerically measuring mean escape times, our model reduction does not rely on this assumption.
In this letter, the diffusive coupling among elements is assumed to be global: i.e., every element interacts with all the others.
We generalize our results to a class of networked systems in the companion article~\jointarticle{}.
As illustrated in Fig.~\ref{fig:trajectories}, stronger couplings lead to higher degrees of synchronization among elements.

In the noise-free case ($D = 0$), the model~\eqref{eq:model.original} exhibits bifurcations regarding $K$~\cite{berglund2007Metastability,ashwin2017Fast}.
Due to the diffusive nature of coupling, there always exist two stable and one unstable uniform steady states: i.e., $\bm{x}_0 = (0, \dots, 0)^\top$, $\bm{x}_1 = (1, \dots, 1)^\top$, and $\bm{x}_r = (r, \dots, r)^\top$, respectively.
In the uncoupled limit ($K = 0$), the system has $3^N$ fixed points.
As $K$ increases, non-uniform fixed points disappear through saddle-node bifurcations, leaving only the three uniform steady states.
At the largest bifurcation point $\rmsub{K}{b}$, a pitchfork bifurcation occurs, where two non-uniform saddles collide the uniform unstable node $\bm{x}_r$, which then becomes a saddle.
The linear stability analysis of $\bm{x}_r$ reveals $\rmsub{K}{b} = r (1 - r)$, which is independent of $N$~\cite{SMderiv}.

We define the mean escape time as follows.
First, the system is initialized to the collective background state $\bm{x}_0$.
Then, the first escape time of element $i$ is defined as
\begin{gather}
  \tau_i \coloneq \inf\Bab{t > 0 \,\Big|\, x_i(t) \geq \xi \text{ given } x_i(0) = 0},
\end{gather}
where $\xi$ is a fixed threshold between the background and active states.
We use $\xi = 0.5$ in the following, but this choice does not affect our results as long as $\xi$ is not close to $r$ or $1$.
As $\tau_i$ is defined for each element, we take an average over elements within a system, and define its expectation to be the mean escape time $\bar{\tau}$:
\begin{gather} \label{eq:met.def}
  \bar{\tau} \coloneq \aab{\frac{1}{N} \sum_{i=1}^N \tau_i}.
\end{gather}

For a one-dimensional gradient system, the following formula for the mean first passage time~\cite[Section~5.5]{gardiner2009Stochastic} is known.
Specifically, consider a stochastic process governed by $\dot{x} = -V'(x) + \sqrt{2D} \xi$ on $(-\infty, \xi)$, with a reflecting boundary at $x = -\infty$ and an absorbing boundary at $x = \xi$.
Here, $V(x)$ denotes the associated potential.
The expected time until a particle initially at $x = 0$ exits the interval through $x = \xi$, referred to as the mean first passage time $T$, is given by
\begin{gather} \label{eq:mfpt}
  T = \frac{1}{D} \int_0^\xi \d y \int_{-\infty}^y \d z \exp\pab{-\frac{V(y) - V(z)}{D}}.
\end{gather}

Let $U(x)$ denote the double-well potential satisfying $f(x) = -\d U / \d x$.
In the uncoupled limit ($K = 0$), the mean escape time of a population coincides with that of a single bistable element,
\begin{gather} \label{eq:T0}
  T_0 \coloneq \frac{1}{D} \int_0^\xi \d y \int_{-\infty}^y \d z \exp\pab{-\frac{U(y) - U(z)}{D}}.
\end{gather}
This formula of $T_0$ simplifies to the Eyring-Kramers law~\cite{berglund2013Kramers} in the low-noise limit~\cite{ishii2024Diffusivea}.

\section{Effective one-dimensional dynamics}

To identify dominant driving factors of collective escape, we develop a model-reduction approach to obtain effective one-dimensional dynamics.
First, we consider a case where coupling is weak enough to assume independence among elements~\cite{SMderiv}.
The $N$-variate Fokker-Planck equation (FPE) corresponding to our model [Eq.~\eqref{eq:model.original}] governs the evolution of the $N$-variate probability density function (PDF), $p_N(x_1, \dots, x_N, t)$.
By marginalizing $p_N$ regarding $x_2, \dots, x_N$, the $N$-variate FPE becomes the evolution equation of the univariate PDF $p_1(x_1, t)$, which depends on the bivariate PDF $p_2(x_1, x_m, t)$ ($m \neq 1$).
To resolve such dependency among PDFs, which is known as BBGKY hierarchy, it is common to introduce the molecular chaos assumption, ignoring correlations among elements.
More precisely, by assuming $p_2(x_i, x_j, t) \approx p_1(x_i, t) p_1(x_j, t)$, one obtains a closed evolution equation of $p_1(x, t)$ as follows:
\begin{gather} \label{eq:nlinmffpe}
  \partial_t p_1 = -\partial_x \bab{f(x) + K \pab{\aab{x} - x}} p_1 + D \partial_x^2 p_1,
\end{gather}
where $\aab{x} = \int x p_1(x, t) \d x$ denotes the expectation of $p_1$.
The derived equation~\eqref{eq:nlinmffpe} involves the mean field $\aab{x}$ and hence nonlinear in $p_1$.
Accordingly, it is called a nonlinear mean-field Fokker--Planck equation (NlinMFFPE).

To calculate mean escape times based on NlinMFFPE, let us introduce the probability current
\begin{gather}
  J(x, t) \coloneq \bab{f(x) + K \pab{\aab{x} - x} - D \partial_x} p_1(x, t),
\end{gather}
with which NlinMFFPE is written as $\partial_t p_1 = -\partial_x J$.
The probability density that passes through $x = \xi$ at time $t$ is $J(\xi, t)$.
Due to the choice of $r = 0.05 \ll 0.5$, the density is expected to always flow towards the active state, implying $J(\xi, t) \geq 0$.
Therefore, one may regard $J(\xi, t)$ as the PDF of the escape time $t$, allowing us to estimate the mean escape time $\bar{\tau}$ by
\begin{gather} \label{eq:met.nlinmffpe}
  \bar{\tau} = \int_0^\infty t \, J(\xi, t) \d t.
\end{gather}
To numerically evaluate Eq.~\eqref{eq:met.nlinmffpe}, we first solve NlinMFFPE [Eq.~\eqref{eq:nlinmffpe}] to obtain time series of $J(\xi, t)$, and then approximate the integral by a sum.

Next, we consider the strong-coupling regime.
Introducing the mean field
\begin{gather}
  X \coloneq \frac{1}{N} \sum_{i=1}^N x_i,
\end{gather}
one can rewrite the model [Eq.~\eqref{eq:model.original}] as:
\begin{gather} \label{eq:model.X}
  \dot{x}_i = f(x_i) + K \pab{X - x_i} + \sqrt{2D} \xi_i.
\end{gather}
Let $y_i$ denote the deviation of $x_i$ from the mean field: i.e., $y_i \coloneq x_i - X$.
Because the coupling is diffusive, $y_i$ decreases as $K$ increases.
Expanding the local bistable flow $f(x_i)$ around $X$, one obtains the following evolution equations of $X$ and $y_i$~\cite{SMderiv}:
\begin{subequations}\begin{align}
  \label{eq:X.dyn}
  \dot{X} =& f(X) + \frac{f''(X)}{2} Z + \sqrt{\frac{2D}{N}} \eta_X,  \\
  \label{eq:Z.def}
  Z \coloneq & \frac{1}{N} \sum_{i=1}^N y_i^2,  \\
  \label{eq:yi.dyn}
  \dot{y}_i =& -\bab{K - f'(X)} y_i + \sqrt{2D} \eta_i,
\end{align}\end{subequations}
where $\eta_X$ and $\eta_i$ are effective zero-mean white Gaussian noise satisfying
\begin{subequations} \begin{align}
 \aab{\eta_X(t) \eta_X(t + \tau)} =& \delta(\tau),  \\
 \label{eq:eta_i.corr}
 \aab{\eta_i(t) \eta_j(t + \tau)} =& \delta(\tau) \pab{\delta_{ij} - \frac{1}{N}},  \\
 \aab{\eta_X(t) \eta_i(t + \tau)} =& 0.
\end{align} \end{subequations}
A notable feature of $X$ dynamics [Eq.~\eqref{eq:X.dyn}] is the appearance of $Z$, the variance within the system, which results from noise.
For sufficiently large $K$ [i.e., $K \gg f'(X)$], $y_i$ dynamics [Eq.~\eqref{eq:yi.dyn}] are independent of $X$, describing an $N$-variate Ornstein-Uhlenbeck (OU) process.
In addition, the correlation function of $\eta_i$ [Eq.~\eqref{eq:eta_i.corr}] suggests that cross-correlations among $\eta_i$ would be negligible for large $N$.
Hence for large $K$ and $N$, each $y_i$ approximately follows a one-dimensional OU process governed by
\begin{gather}
  \dot{y} = -K y + \sqrt{2D} \eta,
\end{gather}
whose stationary variance is $D / K$.
At the same time, when $K$ is large, the time scale of $y_i$ would be much faster than that of $X$.
Hence, approximating as $Z \approx D / K$, one obtains the closed evolution equation of $X$:
\begin{gather} \label{eq:smfd}
  \dot{X} = f(X) + \frac{f''(X)}{2} \frac{D}{K} + \sqrt{\frac{2D}{N}} \eta_X,
\end{gather}
which we call the stochastic mean-field dynamics (SMFD), following a previous work~\cite{dierkes2012Meanfield}.
Its central assumption is that $x_i$ follows a Gaussian distribution around $X$, which is implied by the OU approximation of $y_i$.
The contribution of the variance $Z$ to the dynamics originates from the system's nonlinearity, reflected in non-vanishing $f''(X)$.
As such, the second term of SMFD [Eq.~\eqref{eq:smfd}] clearly illustrates the synergistic effect of nonlinearity, coupling with strength $K$, and noise with intensity $D$.
SMFD can be written using the asspciated potential. 
Thus, the mean first passage time formula [Eq.~\eqref{eq:mfpt}] allows us to calculate mean escape times.
In the strong-coupling limit ($K \to \infty$), SMFD simplifies to $\dot{X} = f(X) + \sqrt{2D/N} \eta_X$, based on which the asymptotic value of the mean escape time, $T_\infty(N)$, is obtained.
The limiting dynamics illustrate the reduction in the effective noise strength by $1 / N$ due to diffusive coupling.

\begin{figure}[t]
  \centering
  \includegraphics[width=\linewidth]{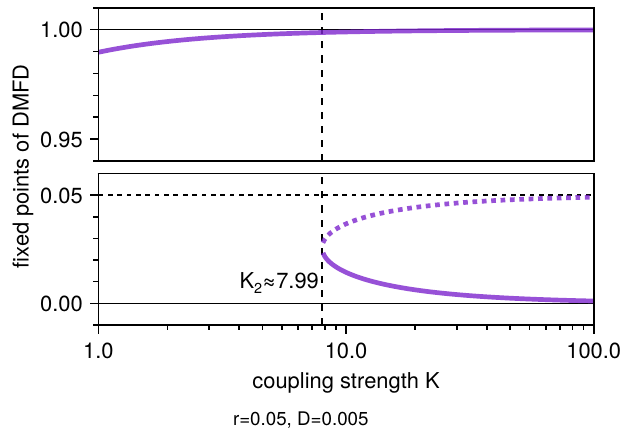}
  \caption{Bifurcation diagram of deterministic mean-field dynamics (DMFD).
    Solid and dashed lines denote stable and unstable branches, respectively.
    The system is monostable for $K < K_2 \approx 7.99$ and bistable otherwise.
  }
  \label{fig:dmfd.bd}
\end{figure}

For sufficiently large $N$, the diffusion term of SMFD [Eq.~\eqref{eq:smfd}] is negligible, yielding the following deterministic mean-field dynamics (DMFD):
\begin{gather} \label{eq:dmfd}
  \dot{X} 
    = g(X; K)
    \coloneq f(X) + \frac{f''(X)}{2} \frac{D}{K}.
\end{gather}
As $K$ is varied, this system exhibits a saddle-node bifurcation at $K = K_2$ as shown in Fig.~\ref{fig:dmfd.bd}.
In particular, the system is monostable for $K < K_2$ with a unique sink corresponding to the active state ($X \approx 1$).
Put differently, the escape of the mean field --- and hence all elements --- occurs deterministically in this intermediate regime.
Here, the influence of random forcing on the mean field is negligible: the noise-induced diversity within the system is what drives the collective escape.

One can also derive DMFD from NlinMFFPE under the same set of assumptions: i.e., large $N$ and normally distributed $y$.
For large $N$, one may assume the equivalence of the mean field $X(t)$ and the expectation $\aab{x}(t)$.
Then, the evolution of $X$ is governed by
\begin{gather} \label{eq:expect.dyn}
  \dot{X} = \odv{}{t} \int x p_1(x, t) \d x = \int x \partial_t p_1(x, t) \d x,
\end{gather}
where one can employ NlinMFFPE [Eq.~\eqref{eq:nlinmffpe}] to rewrite $\partial_t p_1$.
Expanding $f(x)$ around $X$ and assuming a Gaussian distribution of $y$ to neglect its third moment $\aab{y^3}$, Eq.~\eqref{eq:expect.dyn} coincides with DMFD [Eq.~\eqref{eq:dmfd}]~\cite{SMderiv}.

Beyond the bifurcation point ($K > K_2$), the stable branch around the background state prohibits a deterministic escape of $X$.
In the monostable regime ($K < K_2$), the mean escape time corresponds to the travel time of $X$ from $X = 0$ to $\xi$, which is calculated by
\begin{gather} \label{eq:met.dmfd}
  \bar{\tau} = \int_{t(X=0)}^{t(X=\xi)} \d t' = \int_0^\xi \frac{\d t}{\d X} \d X = \int_0^\xi \frac{\d X}{g(X; K)}.
\end{gather}

\begin{figure}[t]
  \centering
  \includegraphics[width=\linewidth]{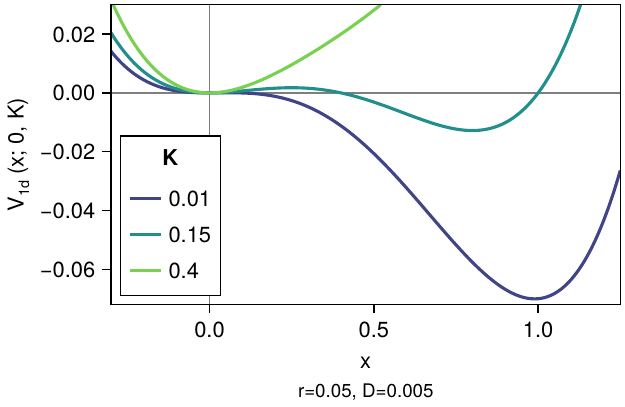}
  \caption{The effective potential $\rmsub{V}{1d}(x; X, K)$ for $X = 0$ at different values of $K$.
    While $\rmsub{V}{1d}$ has two minima when $K$ is small, the influence of diffusive coupling becomes more pronounced for larger $K$, eventually making the system monostable.
  }
  \label{fig:V1d}
\end{figure}

So far, we obtained three reduced descriptions of escape dynamics: NlinMFFPE in the weak-coupling regime, DMFD in the intermediate regime, and SMFD in the strong-coupling regime.
We now determine the regime boundaries of coupling strength $K$ based on properties of the reduced dynamics.
Firstly, we estimate the critical coupling strength $K_1$ above which SMFD is valid.
We choose $K_1$ to be the boundary between the weak-coupling and intermediate regimes, because DMFD is valid for large $N$ as long as SMFD is valid.
The central assumption in deriving SMFD is that $x$ follows a Gaussian distribution around $X$.
To determine $K_1$, let us assume large $N$ to regard $X$ as an external time-dependent parameter independent of $x_i$.
Then, the model [Eq.~\eqref{eq:model.X}] is a one-dimensional system with an associated potential $\rmsub{V}{1d}(x; X, K)$ satisfying $-\d \rmsub{V}{1d} / \d x = f(x) + K \pab{X - x}$.
As illustrated in Fig.~\ref{fig:V1d}, $\rmsub{V}{1d}$ has two minima when $K$ is small, but the influence of diffusive coupling becomes more pronounced as $K$ increases, making $\rmsub{V}{1d}$ closer to a quadratic potential.
Since a quadratic potential, which produces a Gaussian distribution, has no inflection point, SMFD is unlikely to be valid as long as $\rmsub{V}{1d}$ has an inflection point.
Solving $\rmsub{V}{1d}'' = 0$ for $x$~\cite{SMderiv}, one finds that inflection points disappear at
\begin{gather}
  K_1 = \frac{1 - r + r^2}{3}.
\end{gather}
We adopt this value as the boundary $K_1$, because $K > K_1$ should be a necessary condition for the validity of SMFD.
The boundary between the intermediate and strong-coupling regimes, $K_2$, is the saddle-node bifurcation point of DMFD, where DMFD transition from monostable to bistable.
The boundary $K_2$ is significant in that the mean escape time diverges at $K_2$ in the thermodynamic limit ($N \to \infty$).
Put differently, $K_2$ separates phases where the escape rate $1 / \bar{\tau}$ is finite ($K < K_2$) and zero ($K > K_2$) in this limit.
The bifurcation point $K_2$ can be found numerically by solving $g(X_0; K) = 0$ for $K$, where $X_0$ is the position of the minimum of $g(X; K)$~\cite{SMderiv}.
For the current parameter values ($r = 0.05$ and $D = 0.005$), we obtained $K_2 \approx 7.99$.

\section{Numerical verification}
\begin{figure*}[t]
  \centering
  \includegraphics[width=\linewidth]{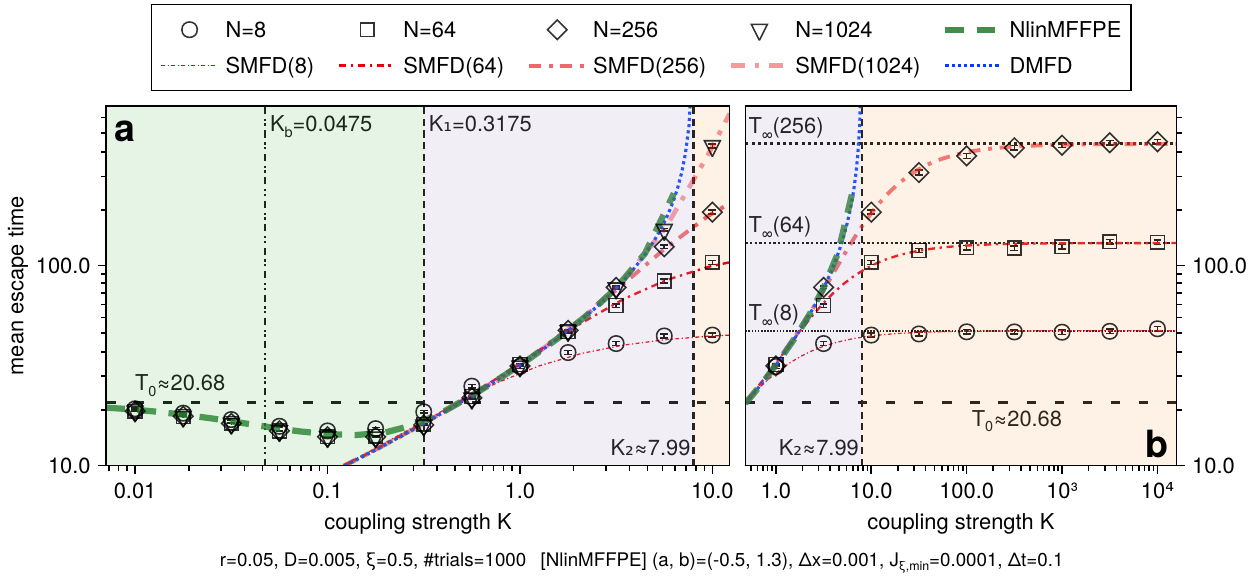}
  \caption{Numerically measured mean escape times (markers) and theoretical predictions (lines) as a function of coupling strength $K$.
    The numerical results agreed well with the theoretical predictions.
    Background colors distinguish the three regimes (weak-coupling: green; intermediate: purple; and strong-coupling: yellow), separated by $K_1$ and $K_2$.
    The horizontal dashed line indicates the uncoupled mean escape time $T_0$ [Eq.~\eqref{eq:T0}].
    \panlab{a} Weak-coupling and intermediate regimes.
    \panlab{b} Intermediate and strong-coupling regimes. The case $N = 1024$ was omitted. 
      Strong-coupling limits of mean escape times $T_\infty(N)$, obtained by replacing $D$ with $D / N$ in Eq.~\eqref{eq:T0}, are shown by the horizontal dotted lines.
  }
  \label{fig:met.vs.K}
\end{figure*}

Thus far, we obtained the following theoretical predictions.
First, in the weak-coupling and intermediate regimes ($K < K_2$), the mean escape time is estimated by Eq.~\eqref{eq:met.nlinmffpe} based on NlinMFFPE [Eq.~\eqref{eq:nlinmffpe}].
Second, in the intermediate and strong-coupling regimes ($K > K_1$), the mean first passage time formula [Eq.~\eqref{eq:mfpt}] with the effective potential and noise strength ($D / N$) of SMFD [Eq.~\eqref{eq:smfd}] predicts mean escape times.
Third, in the intermediate regime, Eq.~\eqref{eq:met.dmfd} based on DMFD [Eq.~\eqref{eq:dmfd}] is also capable of estimating mean escape times.
To verify them, we numerically solved the model SDEs and measured mean escape times~\cite{SMnum}.
The result is shown in Fig.~\ref{fig:met.vs.K}, where measured and predicted mean escape times are compared.
Overall, the theoretical lines agreed well with the numerical results (markers).
As expected, only NlinMFFPE could reproduce the numerical result in the weak-coupling regime; SMFD came to be valid around $K_1$; and NlinMFFPE and DMFD coincided in the intermediate regime between $K_1$ and $K_2$.
Predictions of SMFD deviated from those of DMFD at smaller $K$ for smaller $N$, because the fluctuation in $X$ is stronger for smaller systems, invalidating the deterministic approximation.
We also note that the relation between the mean escape time and $K$ was non-monotonic beyond the largest bifurcation point $\rmsub{K}{b}$, which cannot be explained by bifurcations.

\section{Discussion}
To analyze noise-induced escape in fully connected populations of bistable elements, we derived effective one-dimensional dynamics in three regimes of coupling strength $K$.
In the weak-coupling regime below $K_1$, we obtained nonlinear mean-field Fokker-Planck equation (NlinMFFPE) [Eq.~\eqref{eq:nlinmffpe}].
In the strong-coupling regime beyond $K_2$, stochastic mean-field dynamics (SMFD) [Eq.~\eqref{eq:smfd}] was derived.
In the intermediate regime between $K_1$ and $K_2$, both NlinMFFPE and SMFD reduced to deterministic mean-field dynamics (DMFD) [Eq.~\eqref{eq:dmfd}].
The population is synchronized when $K > K_1$, in the sense that one can characterize the collective escape only by tracking the mean field.
In contrast, escapes are non-synchronous in the weak-coupling regime, as the distribution $p_1$ is necessary for describing the collective dynamics.
Among the two synchronized (i.e., intermediate and strong-coupling) regimes, whereas the fluctuation of the mean field is crucial for the collective escape in the strong-coupling regime, only the variance within the population is sufficient in the intermediate regime.
In short, we identified three qualitatively distinct mechanisms of collective noise-induced escape.
Importantly, all mechanisms involve the interplay of nonlinearity, diffusive coupling, and dynamical noise.

We emphasize that, although both previous studies and our research have divided the parameter space into three regimes, our classification is different from the previous one based on bifurcations.
Indeed, our weak-coupling regime spans all --- i.e., weak-coupling, slow-domino, and fast-domino --- regimes proposed in the literature~\cite{berglund2007Metastability,ashwin2017Fast}, as seen from $\rmsub{K}{b} < K_1$, where $\rmsub{K}{b}$ is the largest bifurcation point.
Interestingly, in the marginal region between $\rmsub{K}{b}$ and $K_1$, escapes are non-synchronous in the sense discussed above, even though non-uniform steady states no longer exist.

The following two assumptions are vital for our results.
The first is the diffusive nature of the coupling.
While our results may qualitatively apply to cases of nonlinear diffusive coupling, it would be a nontrivial task to generalize our approach to non-diffusive cases.
The other is the homogeneity of the isolated dynamics of each element.
If, for instance, values of $r$ and $D$ had significantly differed among elements, it would have been much more challenging to derive low-dimensional reduced descriptions.

Our analysis can be generalized in several aspects.
Firstly, we expect that our results are insensitive to the definition of mean escape times.
This insensitivity would be especially true in the intermediate and strong-coupling regimes, where all the elements are synchronized.
Indeed, we estimated the mean escape time in these regimes by calculating that of the mean field, which technically differed from our mean escape time $\bar{\tau}$ [Eq.~\eqref{eq:met.def}].
Secondly, while the assumption of global coupling greatly simplified the analysis in this letter, it is possible to extend our approach to a class of networked systems; the extension is presented in the companion article~\jointarticle{}.
It turns out that the insights of this letter remain qualitatively valid in the networked systems.
Thirdly, the SMFD framework developed in this research is applicable to other diffusively-coupled nonlinear systems with white Gaussian noise.
The framework was inspired by the study of a model of elastically coupled hair bundles~\cite{dierkes2012Meanfield}, which focused on the strong-coupling limit.
We refined the SMFD formulation to obtain an effective one-dimensional description at finite coupling strength in a more general context.
A further search for synergistic phenomena arising from the interplay of nonlinearity, diffusive coupling, and dynamical noise would be a stimulating direction for future research.

\begin{acknowledgments}
H.I. acknowledges support from the World-leading Innovative Graduate Study Program in Proactive Environmental Studies (WINGS-PES), the University of Tokyo, and JSPS KAKENHI Grant Number JP24KJ0635.
This study was also supported by the JSPS Core-to-Core Program ``Advanced core-to-core network for the physics of self-organizing active matter'' JPJSCCA20230002 to H.I. and H.K.
\end{acknowledgments}

\paragraph{Data Availability.}
The data and code that support the findings of this study are openly available at \url{https://github.com/ishiihidemasa/26-escape-fully-connected}.


\nocite{rackauckas2017Differentialequationsjla,tsitouras2011Runge,rossler2010Runge,rackauckas2017Adaptive,danisch2021Makiejl}

\bibliography{25-phd-thesis_all2all}

\end{document}


\maketitle
\captionsetup{width=.9\linewidth}

\section{Linear stability analysis of the uniform unstable steady state}

We calculate the largest bifurcation point of the system, which is associated with the uniform unstable steady state $\bm{x}_r = \pab{r, \dots, r}^\top$.
In the simplest case of $N = 2$, the Jacobian matrix of the system is
\begin{gather}
  J_2\pab{\bm{x}} = \begin{pmatrix}
    f'(x_1) - \frac{K}{2} & \frac{K}{2}  \\
    \frac{K}{2} & f'(x_2) - \frac{K}{2}
  \end{pmatrix},
\end{gather}
where $f'(x) = \d f / \d x$.
At $\bm{x}_r$, the eigenvalues of $J_2(\bm{x}_r)$ are $\lambda_+ = f'(r)$ and $\lambda_- = f'(r) - K$, whose corresponding eigenvectors are $\pab{1, 1}^\top$ and $\pab{1, -1}^\top$, respectively.
The eigenvalue $\lambda_+$ indicates the growth rate of a uniform perturbation along the diagonal line in the phase space.
Since $f'(r) = r \pab{1 - r} > 0$, the fixed point $\bm{x}_r$ is repelling in this uniform mode.
The stability of the other mode in the direction of $\pab{1, -1}^\top$ depends on the coupling strength $K$.
When $K < f'(r)$, the eigenvalue $\lambda_-$ is positive, and hence the fixed point $\bm{x}_r$ is a source.
Otherwise --- when $K > f'(r)$ --- the eigenvalue $\lambda_-$ is negative, making $\bm{x}_r$ a saddle.
The bifurcation point, where a source turns into a saddle, is $\rmsub{K}{b} = f'(r) = r \pab{1 - r}$.

In the general $N$ element cases, the eigenvalues of the Jacobian matrix at $\bm{x}_r$ remain the same.
The Jacobian matrix is
\begin{gather}
  J_N(\bm{x}) = \begin{pmatrix}
    f'(x_1) - \frac{K \pab{N - 1}}{N} & K / N & \cdots & K / N \\
    K / N & f'(x_2) - \frac{K \pab{N - 1}}{N} &  & \vdots \\
    \vdots &  & \ddots & K / N \\
    K / N & \cdots & K / N & f'(x_N) - \frac{K \pab{N - 1}}{N}
  \end{pmatrix}.
\end{gather}
At the unstable uniform fixed point $\bm{x}_r$, a uniform perturbation $\bm{\delta}_+ = \pab{1, \dots, 1}^\top$ satisfies
\begin{gather}
  J_N(\bm{x}_r) \, \bm{\delta}_+ = f'(r) \, \bm{\delta}_+,
\end{gather}
indicating that $\bm{\delta}_+$ is an eigenvector with the eigenvalue $\lambda_+ = f'(r)$.
Inspired by the other eigenvector $\pab{1, -1}^\top$ in the $N = 2$ case, let us consider a non-uniform perturbation of the form $\bm{\delta}^{(2)}_- = \pab{1, -1, 0, \dots}^\top$.
It satisfies
\begin{gather}
  J_N(\bm{x}_r) \, \bm{\delta}^{(2)}_- = \bab{f'(r) - K} \, \bm{\delta}^{(2)}_-,
\end{gather}
indicating that $\bm{\delta}^{(2)}_-$ is another eigenvector of $J_N(\bm{x}_r)$ with the eigenvalue of $\lambda_- = f'(r) - K$.
More generally, one can consider a non-uniform perturbation $\bm{\delta}^{(i)}_-$, whose first component is $1$, $i$th component is $-1$, and all the others are $0$.
Due to the symmetry of the model, all the $N - 1$ vectors $\bm{\delta}^{(i)}_-$ are also eigenvectors with the same eigenvalue $\lambda_-$.
Considering that $\bm{\delta}^{(i)}_-$ are linearly independent, $\lambda_-$ is the eigenvalue of $J_N(\bm{x}_r)$ with the multiplicity of $N - 1$.
Since the rank of $J_N(\bm{x}_r)$ is $N$, this establishes that an eigenvalue of $J_N(\bm{x}_r)$ is either $\lambda_+$ or $\lambda_-$.
Consequently, the bifurcation point $\rmsub{K}{b} = f'(r)$, where $\lambda_-$ equals zero, does not depend on the system size $N$.

\section{Derivation of nonlinear mean-field Fokker--Planck equation}

The $N$-variate Fokker--Planck equation (FPE) corresponding to our model [Eq.~\maneqref{1}] reads
\begin{gather} \label{eq:FPE.ndim}
  \partial_t p_N = -\sum_{j=1}^{N} \partial_{x_j} \pab{\bab{f(x_j) + \frac{K}{N} \sum_{m=1}^{N} \pab{x_m - x_j}} p_N} + D \sum_{i, j} \partial_{x_i x_j} p_N,
\end{gather}
where $p_N(x_1, \dots, x_N, t)$ denotes the $N$-particle probability distribution function (PDF).
To derive the nonlinear mean-field FPE (NlinMFFPE), we first integrate the FPE [Eq.~\eqref{eq:FPE.ndim}] over all variables except $x_1$.
By this marginalization, the left-hand side (LHS) simply becomes
\begin{gather}
  \partial_t p_1(x_1, t).
\end{gather}
In the right-hand side (RHS), we begin with the drift terms.
We introduce
\begin{gather}
  A_0(x) \coloneq f(x) - K x
\end{gather}
to simplify the notation.
Within the drift terms, the following part,
\begin{gather}
  \int \prod_{l=2}^{N} \d x_l \sum_{j=1}^{N} \partial_{x_j} \bab{A_0(x_j) p_N},
\end{gather}
does not involve interaction terms.
We first calculate the case of $j = 1$ within the summation:
\begin{subequations} \begin{align}
  \int \prod_{l=2}^{N} \d x_l \partial_{x_1} \bab{A_0(x_1) p_N}
    =& \int \prod_{l=2}^{N} \d x_l \bab{p_N \partial_{x_1} A_0(x_1) + A_0(x_1) \partial_{x_1} p_N}  \\
    =& p_1(x_1, t) \partial_{x_1} A_0(x_1) + A_0(x_1) \partial_{x_1} p_1(x_1, t)  \\
    =& \partial_{x_1} \bab{A_0(x_1) p_1(x_1, t)}.
\end{align} \end{subequations}
To calculate the other terms (i.e., $j \neq 1$), it suffices to consider the case of $j = 2$:
\begin{subequations} \begin{align} \notag
  & \int \prod_{l=2}^{N} \d x_l \partial_{x_2} \bab{A_0(x_2) p_N}  \\
  =& \int \d x_2 \prod_{l=3}^{N} \d x_l \bab{p_N \partial_{x_2} A_0(x_2) + A_0(x_2) \partial_{x_2} p_N}  \\
  =& \int \d x_2 \bab{p_2(x_1, x_2, t) \partial_{x_2} A_0(x_2) + A_0(x_2) \partial_{x_2} p_2(x_1, x_2, t)}  \\
  =& \int \d x_2 \partial_{x_2} \bab{A_0(x_2) p_2(x_1, x_2, t)}  \\
  =& \bab[bigg]{A_0(x_2) p_2(x_1, x_2, t)}_{x_2=-\infty}^{x_2=\infty}  \\
  =& 0.
\end{align} \end{subequations}
To obtain the last equality, we dropped the surface term, assuming $p_2(x_1, \pm\infty, t) \equiv 0$.
Similar assumptions will be made in the following.
The interaction terms can be rewritten as
\begin{subequations} \begin{align}
  \int \prod_{l=2}^{N} \d x_l \partial_{x_1} \bab{\frac{K}{N} \sum_{m=1}^{N} x_m p_N}
    =& \frac{K}{N} \int \prod_{l=2}^{N} \d x_l \sum_{m=1}^{N} \partial_{x_1} \bab{x_m p_N}  \\
    =& \frac{K}{N} \int \prod_{l=2}^{N} \d x_l \bab{
      \partial_{x_1} \pab{x_1 p_N} + \sum_{m=2}^{N} x_m \partial_{x_1} p_N
    }.
\end{align} \end{subequations}
The first term with $x_1$ becomes
\begin{gather}
  \frac{K}{N} \int \prod_{l=2}^{N} \d x_l \partial_{x_1} \pab{x_1 p_N}
  = \frac{K}{N} \partial_{x_1} \bab{x_1 p_1(x_1, t)}.
\end{gather}
To calculate the second part involving the sum, we first consider the case of $m = 2$:
\begin{subequations} \begin{align}
  \int \prod_{l=2}^{N} \d x_l x_2 \partial_{x_1} p_N
  =& \int \d x_2 x_2 \int \prod_{l=3}^{N} \d x_l \partial_{x_1} p_N  \\
  =& \int \d x_2 x_2 \partial_{x_1} p_2(x_1, x_2, t)  \\
  =& \partial_{x_1} \int \d x_2 x_2 p_2(x_1, x_2, t).
\end{align} \end{subequations}
It follows that
\begin{gather}
  \sum_{m=2}^{N} \int \prod_{l=2}^{N} \d x_l x_m \partial_{x_1} p_N
    = \partial_{x_1} \sum_{m=2}^{N} \int \d x_m x_m p_2(x_1, x_m, t).
\end{gather}
To summarize, marginalizing the drift terms resulted in
\begin{gather}
  -\partial_{x_1} \bab{A_0(x_1) p_1 + \frac{K}{N} x_1 p_1 + \sum_{m=2}^{N} \int \d x_m x_m \, p_2(x_1, x_m, t)}.
\end{gather}
Let us now calculate the diffusion terms, which can be rewritten as
\begin{gather} \begin{aligned}
  & \int \prod_{l=2}^{N} \d x_l \sum_{i, j} \partial_{x_i} \partial_{x_j} p_N  \\
  =& \partial_{x_1}^2 p_1(x_1, t) 
    + 2 \partial_{x_1} \sum_{j=2}^{N} \int \prod_{l=2}^{N} \d x_l \partial_{x_j} p_N
    + \sum_{i,j=2}^{N} \int \prod_{l=2}^{N} \d x_l \partial_{x_i} \partial_{x_j} p_N.
\end{aligned} \end{gather}
To calculate the second part, we consider the case of $j = 2$:
\begin{subequations} \begin{align}
  \partial_{x_1} \int \d x_2 \int \prod_{l=3}^{N} \d x_l \partial_{x_2} p_N
  =& \partial_{x_1} \int \d x_2 \partial_{x_2} p_2(x_1, x_2, t)  \\
  =& \partial_{x_1} \bab[bigg]{p_2(x_1, x_2, t)}_{x_2=-\infty}^{x_2=\infty}  \\
  =& 0.
\end{align} \end{subequations}
The same holds for $j \geq 3$, and hence the second part equals zero.
The last part also vanishes:
\begin{subequations} \begin{align}
  \sum_{i,j=2}^{N} \int \d x_i \d x_j \prod_{l \geq 2, l \neq i, j} \d x_l \partial_{x_i x_j} p_N
  =& \sum_{i,j=2}^{N} \iint \d x_i \d x_j \partial_{x_i} \partial_{x_j} p_3(x_1, x_i, x_j, t)  \\
  =& \sum_{i,j=2}^{N} \int \d x_i \partial_{x_i} \bab[bigg]{p_3(x_1, x_i, x_j, t)}_{x_j=-\infty}^{x_j=\infty}  \\
  =& 0.
\end{align} \end{subequations}
Put together, the diffusion terms simplifies through marginalization to
\begin{gather}
  D \partial_{x_1}^2 p_1(x_1, t).
\end{gather}
Combining the drift and diffusion terms, one obtains the following marginalized FPE for $p_1(x_1, t)$:
\begin{gather} \label{eq:p1.fpe.exact}
  \partial_t p_1 = -\partial_{x_1} \bab{\pab[bigg]{f(x_1) - K x_1} p_1 + \frac{K}{N} \pab{x_1 p_1 + \sum_{m=2}^{N} \int \d x_m \, x_m p_2(x_1, x_m, t)}} + D \partial_{x_1}^2 p_1.
\end{gather}

One sees that the evolution of a univariate PDF $p_1(x_1, t)$ depends on the bivariate PDFs $p_2(x_1, x_m, t)$ ($m \neq 1$).
Such hierarchical dependence among PDFs is known as the BBGKY hierarchy.
To resolve this dependency, the molecular chaos assumption is introduced:
\begin{gather}
  p_2(x_i, x_j, t) \approx p_1(x_i, t) \, p_1(x_j, t).
\end{gather}
This entails two sub-assumptions.
The first is the independence among elements, which justifies expressing the bivariate PDF as a product of two univariate PDFs.
The second is the indistinguishability.
In general, the functional form of $p_1$ may differ across elements: i.e., $p_1$ has indices of elements as $p_1^{(i)}(x_i, t)$.
The molecular chaos assumption implies the sub-assumption of $p_1^{(i)}(x_i, t) \equiv p_1(x, t)$ for all $i$.
In the current setting, all elements are indeed indistinguishable.
Therefore, the validity of the assumption is contingent on the independence among elements.
One expects that as the coupling becomes stronger, the independence assumption gets less valid.

Under the molecular chaos assumption, the integral involving $p_2$ can be simplified in the following manner:
\begin{subequations} \begin{align}
  \frac1N \sum_{m=2}^{N} \int \d x_m \, x_m p_2(x_1, x_m, t)
  \approx& \frac1N \sum_{m=2}^{N} \int \d x_m \, x_m p_1(x_1, t) p_1(x_m, t)  \\
  =& p_1(x_1, t) \frac{N - 1}{N} \int \d x \, x \, p_1(x, t)  \\
  =& \frac{N - 1}{N} \aab{x} p_1(x_1, t),
\end{align} \end{subequations}
where $\aab{x}$ is the expectation regarding $p_1(x, t)$:
\begin{gather}
  \aab{x} = \int_{-\infty}^\infty x p_1(x, t) \d x.
\end{gather}
Therefore, the evolution equation of $p_1(x, t)$ for large $N$ reads
\begin{gather} \label{eq:nlinmffpe}
  \partial_t p_1 = -\partial_x \Bab{\bab{f(x) + K \pab[big]{\aab{x} - x}} p_1} + D \partial_x^2 p_1,
\end{gather}
which is Eq.~\maneqref{6} in the main text.

\section{Evolution of mean field and deviations}

Assuming that deviations $y_i \coloneq x_i - X$ are small, we expand $f(x)$ as
\begin{gather}
  f(x) = f(X) + f'(X) y + \frac{f''(X)}{2} y^2 + \frac{f'''(X)}{6} y^3 + \order{y^4}.
\end{gather}
Noting that $\sum_{i=1}^N y_i = 0$, the evolution of the mean field is by definition
\begin{gather} \label{eq:X.dyn}
  \dot{X} = \frac{1}{N} \sum_{i=1}^N \dot{x}_i
    \approx f(X) + \frac{f''(X)}{2} \underbrace{\frac{1}{N} \sum_{i=1}^N y_i^2}_{\eqcolon Z} + \frac{f'''(X)}{6} \frac{1}{N} \sum_{i=1}^N y_i^3 + \sqrt{2D} \hat{\eta}_X,
\end{gather}
where we introduced
\begin{gather}
  \hat{\eta}_X \coloneq \frac{1}{N} \sum_{i=1}^N \xi_i.
\end{gather}
It follows from the properties of $\xi_i$ that $\hat{\eta}_X$ is a Gaussian random variable satisfying
\begin{gather}
  \aab{\hat{\eta}_X(t)} = 0,  \qquad
  \aab{\hat{\eta}_X(t) \hat{\eta}_X(t + \tau)} = \frac{1}{N} \delta\pab{\tau}.
\end{gather}
In the main text, we instead introduced
\begin{gather}
  \eta_X \coloneq \frac{1}{\sqrt{N}} \sum_{i=1}^N \xi_i
\end{gather}
to normalize its variance, so that the reduction in the effective noise strength from $D$ to $D / N$ is more visible.
The evolution of deviations is, to the first order,
\begin{gather} \label{eq:yi.dyn}
  \dot{y}_i = \dot{x}_i - \dot{X} \approx -\bab{K - f'(X)} y_i + \sqrt{2D} \eta_i, 
\end{gather}
where
\begin{gather}
  \eta_i \coloneq \xi_i - \frac{1}{N} \sum_{j=1}^N \xi_j = \xi_i - \frac{\eta_X}{\sqrt{N}}.
\end{gather}
Again, it is a Gaussian random variable with the following statistical properties:
\begin{subequations} \begin{gather}
  \aab{\eta_i(t)} = 0,  \\
  \aab{\eta_i(t) \eta_X(t+\tau)} = 0,  \\
  \aab{\eta_i(t) \eta_j(t+\tau)} = \delta\pab{\tau} \pab{\delta_{ij} - \frac{1}{N}}.
\end{gather} \end{subequations}
As $y_i$ dynamics [Eq.~\eqref{eq:yi.dyn}] describe an Ornstein-Uhlenbeck (OU) process, $y_i$ would follow a Gaussian distribution in a stationary state.
Therefore, the third moment of deviations --- i.e., $\aab{y^3} \approx \frac{1}{N} \sum_{i=1}^N y_i^3$ --- would vanish, simplifying $X$ dynamics [Eq.~\eqref{eq:X.dyn}] to Eq.~\maneqref{11a} in the main text.

\section{Derivation of deterministic mean-field dynamics from nonlinear mean-field Fokker-Planck equation}

As explained in the main text, we assume the equivalence between $X(t)$ and $\aab{x}(t)$, which corresponds to assuming large $N$.
The evolution of the mean field is governed by
\begin{subequations} \begin{align}
  \dot{X} =& \int x \partial_t p_1(x, t) \d x  \\
    =& -\int \Bab{x \partial_x \bab{f(x) p_1(x, t)} + K x \bab{\partial_x \pab{\aab{x} - x} p_1(x, t)} + D x \partial_x^2 p_1(x, t)} \d x.
\end{align} \end{subequations}
We calculate each term on the RHS, integrating by parts and dropping surface terms:
\begin{subequations} \begin{gather}
  \int x \partial_x \bab{f(x) p_1(x, t)} \d x 
    = \bab{x f(x) p_1(x, t)}_{-\infty}^{\infty} - \int f(x) p_1(x, t) \d x
    = -\aab{f(x)},  \\
  \int x \partial_x \bab{\pab{\aab{x} - x} p_1(x, t)} \d x 
    = \bab{x \pab{\aab{x} - x} p_1(x, t)}_{-\infty}^\infty - \int \pab{\aab{x} - x} p_1(x, t) \d x
    = \aab{x} - \aab{x} = 0,  \\
  \int x \partial_x^2 p_1(x, t) \d x
    = \bab{x \partial_x p_1(x, t)}_{-\infty}^{\infty} - \int \partial_x p_1(x, t) \d x
    = 0.
\end{gather} \end{subequations}
Due to the assumption that $X \approx \aab{x}$, we can calculate as $\aab{f(X)} = \aab{f(\aab{x})} = f(X)$.
Hence, expansion of $f(x)$ around $X$ yields the following:
\begin{gather}
  \aab{f(x)} \approx f(X) + f'(X) \aab{y} + \frac{f''(X)}{2} \aab{y^2} + \frac{f'''(X)}{6} \aab{y^3}.  
\end{gather}
It holds that $\aab{y} = 0$.
Moreover, by additionally assuming Gaussianity of $y$, its third moment $\aab{y^3}$ would vanish, resulting in
\begin{gather}
  \dot{X} = f(X) + \frac{f''(X)}{2} Z,
\end{gather}
which is deterministic mean-field dynamics (DMFD) [Eq.~\maneqref{15} in the main text].

\section{Derivation of regime boundaries of coupling strength}

We defined the boundary between the weak-coupling and intermediate regimes, $K_1$, as the point at which inflection points of the effective one-dimensional potential
\begin{gather}
  \rmsub{V}{1d}(x; X, K) = \frac{1}{4} x^4 - \frac{1 + r}{3} x^3 + \frac{r + K}{2} x^2 - KXx
\end{gather}
disappear.
An inflection point $x^*$ satisfies
\begin{gather}
  \odv[order=2]{}{x} \rmsub{V}{1d} = 3x^2 - 2\pab{1 + r} x + r + K = \pab{x^* - x^*_+} \pab{x^* - x^*_-} = 0,
\end{gather}
where
\begin{gather}
  x^*_{\pm} = \frac{1 + r \pm \sqrt{1 - r + r^2 - 3K}}{3}.
\end{gather}
It follows that
\begin{gather}
  K_1 = \frac{1 - r + r^2}{3}.
\end{gather}

The boundary between the intermediate and strong-coupling regimes, $K_2$, is the bifurcation point of deterministic mean-field dynamics (DMFD) [Eq.~\maneqref{15} in the main text].
To numerically obtain the bifurcation point, we calculate the position of the minimum of DMFD's flow,
\begin{gather}
  g(X; K) = -X^3 + \pab{1 + r} X^2 - \pab{r + \frac{3D}{K}} X + \frac{D \pab{1 + r}}{K},
\end{gather}
which is
\begin{gather}
  X_0 = \frac{1}{3} \pab{1 + r - \sqrt{1 - r + r^2 - \frac{9D}{K}}}.
\end{gather}
For small $K$, $g(X_0; K)$ is positive.
It declines as $K$ increases, and a saddle-node bifurcation occurs when $g(X_0; K) = 0$.
Beyond this bifurcation point $K_2$, DMFD is bistable.
Accordingly, one can determine $K_2$ by numerically solving $g(X_0, K) = 0$ for $K$.

\section{Numerical methods}
All numerical analysis was performed using Julia in Docker containers in addition to Apptainer containers on a high performance computer.
The data and code that support the findings of this study are openly available at \url{https://github.com/ishiihidemasa/26-escape-fully-connected}.
For numerical integration of differential equations, \texttt{DifferentialEquations.jl}~\cite{rackauckas2017Differentialequationsjla} was utilized.
As specific solvers, a variant of Runge-Kutta method, \texttt{Tsit5()}~\cite{tsitouras2011Runge}, was used for ODE systems such as DMFD.
For SDE systems, an adaptive version of a stochastic Runge-Kutta scheme~\cite{rossler2010Runge}, \texttt{SOSRA()}, was used~\cite{rackauckas2017Adaptive}.
An escape of each element, which is an upward zero-crossing event of $x_i - \xi$, was detected using Callback functionalities, particularly \texttt{SciMLBase.ContinuousCallback}.
Figures were generated using \texttt{CairoMakie.jl}, which is a backend of \texttt{Makie.jl}~\cite{danisch2021Makiejl}.

To predict mean escape times based on NlinMFFPE [i.e., using Eq.~\maneqref{8}], we first solved a discretized version of NlinMFFPE in the interval $x \in (a, b)$.
Spatial derivatives were discretized with the step size $\Delta x$ according to the central difference scheme. 
The resulting ODE system was solved using $\texttt{Tsit5()}$.
The numerical integration was terminated when the mean field $X$ approached the active state (specifically, $X > 0.9$) and the probability current at the threshold, $J(\xi, t)$, became small to satisfy $J(\xi, t) \leq J_{\xi,\mathrm{min}}$.
The termination functionality was implemented using Callback functionalities.
Results of numerical integration were recorded with the time interval of $\Delta t$.
Specific values of numerical parameters are shown inside relevant figures so as to eliminate a room for a human error in transcription.
The integral in Eq.~\maneqref{8} was approximated by a sum, which was calculated using the obtained time series of $J(\xi, t)$.
The integral in Eq.~\maneqref{17} was also approximated by a sum.
On the other hand, \texttt{Integrals.jl} was utilized to approximate the integrals in the mean first passage time formula [Eq.~\maneqref{4}].
The bifurcation point of DMFD, $K_2$, was calculated using \texttt{find\_zero()} of \texttt{Roots.jl}.
Specific parameter values are shown in each figure to avoid mistakes in transcription.

\printbibliography